\documentclass[aps,pre,preprint,floatfix]{revtex4-1}
\usepackage{amsmath,amsfonts,amssymb}
\usepackage{graphicx}
\usepackage{dcolumn}
\usepackage{bm}
\usepackage{verbatim}
\usepackage{float}
\usepackage{color}
\graphicspath{{./figures/}}

\begin{document}

\title{Gradient-driven diffusion and pattern formation in crowded mixtures}


\author{Prithviraj Nandigrami}
\affiliation{Department of Physics, Kent State University}
\author{Brandy Grove}
\affiliation{Department of Macromolecular Science and Engineering, Case Western Reserve University}
\author{Andrew Konya}
\affiliation{Liquid Crystal Institute, Kent State University}
\author{Robin L. B. Selinger}
\affiliation{Liquid Crystal Institute, Kent State University}

\date{\today}



\begin{abstract}

\noindent

Gradient-driven diffusion in crowded, multicomponent mixtures is a topic
of high interest because of its role in biological processes such as transport
in cell membranes. In partially phase-separated solutions, gradient-driven diffusion affects microstructure,
which in turn affects diffusivity; a key question is how this complex coupling
controls both transport and pattern formation.
To examine these mechanisms, we study a two-dimensional
multi-component lattice gas model, where ``tracer'' molecules diffuse between 
a source and a sink separated by a solution of sticky ``crowder'' molecules that cluster
to form dynamically evolving obstacles.  In the high temperature limit,
crowders and tracers are miscible and transport may be predicted analytically.  
At intermediate temperatures, crowders phase separate into clusters that drift toward the tracer sink. 
As a result, steady-state tracer diffusivity depends non-monotonically on both 
temperature and crowder density and we observe a variety of complex microstructures.
In the low temperature limit, crowders rapidly aggregate to form obstacles that are kinetically arrested; 
if crowder density is near the percolation threshold, resulting tracer diffusivity shows scaling
behavior with the same scaling exponent as the random resistor network model. 
Though highly idealized, this simple model reveals fundamental mechanisms 
governing coupled gradient-driven diffusion, phase separation, and microstructural evolution in crowded solutions.

\end{abstract}


\maketitle



\section*{Introduction}

Diffusion, aggregation, and coarsening in multi-component mixtures
leads to a rich variety of morphologies in complex
fluids~\cite{gelbart:96}. In crowded environments, diffusive transport
under a driving force often gives rise to complex pattern formation
including striped structures~\cite{schmittmann:98}. Molecular
diffusion in a crowded mixture containing multiple components affects
both chemical reaction kinetics and pattern
formation~\cite{ames:86,zimmerman:93}. Diffusion in crowded
environments often plays a key role in transport within cell
membranes~\cite{skou:65,simons:97}, and in solutions and
cells~\cite{dix:08}. Brownian motion is often observed in cellular
environments such as bacterial cytoplasm~\cite{english:11}, and other
crowded biological environments~\cite{dlugosz:11}.  
For large macromolecules, the dominant transport behavior in such environments
is diffusive in nature~\cite{golding:06,weber:10,english:11}.

In a multi-component mixture, diffusivity of each molecular species depends on the
density of all components present, interaction energy, and
microstructure, that is, pattern formation during phase separation. 
Thus, diffusive transport and microstructure co-evolve and are coupled in general. 
\color{black}
Typically, multi-component simulation studies consider stationary crowders
while the tracers are allowed to diffuse~\cite{saxton:90,berry:11}. In this paper, we consider 
a system where both tracers and crowders are allowed to diffuse, and where crowders
have attractive interactions and may undergo phase separation to form dynamically
evolving mobile obstacles. 

\color{black}
Diffusive transport in crowded environments has been extensively investigated using a
diverse range of simulation techniques, analytic models, as well as
experiments~\cite{fedders:78,nakazato:80,kehr:81,kutner:82,tahir-kheli:83,kutner:83,kutner:83a,kehr:83,
schoen:84,kutner:85,granek:90,granek:90a,rothman:94,brown:98,vasanthi:02,vrljic:02,jeon:11,zaccone:12,sokolov:12,
jeon:16,ghosh:16}.  
\color{black}
In the recent years, several distinct simulation models have given particular emphasis
on modeling anomalous diffusion processes in crowded medium. Of these, on-lattice
simulation models~\cite{saxton:87,saxton:90,saxton:93,saxton:94,saxton:96,saxton:07,
echeveria:07,saxton:08,vilaseca:11} summarize the different regimes for anomalous
diffusion and how the system evolves to the stationary state. Off-lattice simulation
models also have been proposed to describe anomalous diffusion in crowded media
via Brownian Dynamics~\cite{schoneberg:14} as well as by taking into account
hydrodynamic interactions~\cite{ando:10,ghosh:15}. 
\color{black}
The effect of mobile obstacles on Brownian diffusion was recently investigated by Berry and Chat\'e~\cite{berry:14},  
demonstrating that the nature of obstacle diffusion determines whether tracer motion is diffusive or sub-diffusive.
Recent simulation studies have investigated the effect of bimolecular
chemical reactions in confined environments in the presence of crowding species
in solution~\cite{schmit:09}. The resulting reaction rate
shows sensitive dependence on density of the crowding species. The
effect of macromolecular crowding on the collapse of biopolymers has very
recently been investigated and scaling laws have been proposed~\cite{kang:15}.

In this paper, we examine the coupling of multi-species diffusive
transport, phase separation, and pattern formation using the multi-component  lattice
gas model in two dimensions. The driving force for diffusion is
provided by an applied density gradient, with boundary conditions
on the density of a ``tracer'' species 
at the two edges of the cell, rather
than by a constant external driving force as in related models~\cite{schmittmann:98}.
Tracers diffuse through a solvent with
a prescribed density of ``crowder'' species.  Our model demonstrates
that the resulting tracer flux and crowder microstructure both
depend sensitively on crowder density, interaction strength, and temperature. 
\color{black}
This model could describe, for example, recent experiments by Gericke and coworkers~\cite{neumann:16}, where a 
composition gradient in a lipid membrane is maintained using microfluidic methods, and the resulting lipid composition
profile changes when crowder proteins are introduced.
\color{black}
While this highly idealized model presented in this paper does not
take into account hydrodynamic effects or detailed molecular scale
interactions, it reveals several key fundamental mechanisms by which microstructural
evolution and diffusive transport may mutually interact.

\color{black}
\section*{Model}

We perform Monte Carlo (MC) simulations of a two-dimensional lattice
gas model with two diffusing species, ``tracers'', and
``crowders''. The system Hamiltonian is
$H = \sum\limits_{i, \, j} U\Big(s_{i},s_{j}\Big)$, where the sum is over
nearest neighbor pairs and $s_{i}=1,2,3$ refer to vacancy, tracer, and
crowder particle types, respectively. At most one particle may occupy
each lattice site, and vacancies are reminiscent of background solvent. 
Crowders have attractive nearest neighbor
interaction energy 
\color{black}
$U(3,3) = -J_\mathrm{int}$, 
\color{black}
and can thus undergo phase
separation as a function of temperature and density.  All other
interactions are excluded volume only, with
$U(1,1)=U(1,2)=U(1,3)=U(2,1)=U(2,2)=U(2,3)=U(3,1)=U(3,2)=0$. 
Thus, tracers have no energetic preference to aggregate with each other or
bond to crowders.
\color{black}
The dimensionless parameter $k_\mathrm{B}T_\mathrm{s}/J_\mathrm{int}$
sets the energy scale in our model, where $J_\mathrm{int}$ and $k_\mathrm{B}T_\mathrm{s}$
are independent parameters.
\color{black}
The simulation cell is a square lattice of dimension $L$ with periodic
boundary conditions in the y-direction only.  In the initial
state, crowders are randomly distributed throughout the simulation
cell with a density $\rho_{\mathrm{C}}$.  We impose a density gradient
of the tracer species along the x-direction by introducing a
tracer-emitting source along the left side of the cell and a
tracer-absorbing sink along the right side; that is, we impose the
boundary condition that tracer density $\rho_{\mathrm{T}} = 1$ at $x = 0$
and $\rho_{\mathrm{T}} = 0$ at $x=L$.

\color{black}
Both tracer and crowder species diffuse via nearest neighbor hops implemented via
the Metropolis algorithm~\cite{metropolis:53}. 
\color{black}
The particle hopping rule implemented in this model
is analogous to a model of diffusive percolation for ``blind ants'' where
particles hop by choosing a site from all neighboring sites~\cite{selinger:90}.  
Particles may only move into a neighboring
site if it is vacant.  Thus, if there are no vacancies present, the
system arrests and no further diffusion can occur.  As tracers emerge
from the source and are absorbed in the sink, the number of tracers
changes with time while the number of crowders remain constant.
Simulations were run for system size $L=100$ for at least
$2 \times 10^{9}$ Monte Carlo steps, where each Monte Carlo step
represents one attempted move per lattice site. 
\color{black}
Each simulation time step is defined as one attempted
Monte Carlo move per lattice site.
\color{black}
Such long simulations are necessary to allow the system to reach steady state, even for such relatively small system sizes.
\color{black}
We perform ten discrete simulation runs for each point in the phase space
defined by ($\rho_\mathrm{C}$, $T_\mathrm{s}$). Simulated ensemble for each
point in the phase space is constructed by averaging over the last 10\%
data of each simulation to capture the steady state behavior.
\color{black}

We examine the interaction of diffusive transport and microstructural
evolution in this model system.
\color{black}
We assume Fick's law diffusion, where
tracer flux ($J$) is calculated by counting the number of tracer particles annihilating
at the sink per unit length and per unit time; 
Net tracer flux is computed by counting the rate of tracers annihilating at the sink
per unit length per unit time.
We define the net diffusivity of tracers as $D_\mathrm{net}$, and define $D_{0}$ 
as the net diffusivity of tracers in the absence of crowders. 
The quantity $D_\mathrm{net}/D_{0}$ is thus calculated as tracer flux (with crowders) 
normalized by the tracer flux in the absence of crowders.

\color{black}
Crowder particles represent a lattice gas and at low temperature
they phase separate into clusters which gradually coarsen. The presence of tracer
diffusion alters this microstructural evolution: crowders drift toward the sink and form clusters which may
be compact, elongated in the direction of tracer flow, or flattened
against the sink. Likewise, diffusive transport of tracers is
strongly affected by clustering of crowders which represent sticky mobile
obstacles.

In the high temperature limit, where crowder attractive interactions can be neglected, tracer diffusivity
as a function of crowder density can be calculated analytically as
described below. The introduction of attractive interactions
for the crowder species completely changes the system behavior. Even in
this highly idealized model, the resulting tracer diffusivity cannot
be described by a simple function of temperature and crowder density
but shows several characteristic regimes.
Several previous studies have focused on models of diffusion in which
diffusivity is typically reported as a function of crowder density.
In our system, we apply a gradient in tracer diffusivity, which in turn
induces a gradient in crowder density. The resulting pattern formation
process governs the transport of the two species. 
This simplified model thus demonstrates a complex mechanism 
governing gradient-driven transport in multicomponent mixtures.

\section*{Results}
\subsection*{Temporal evolution}

\begin{figure}
  \begin{centering}
    \includegraphics[width=8.75cm]{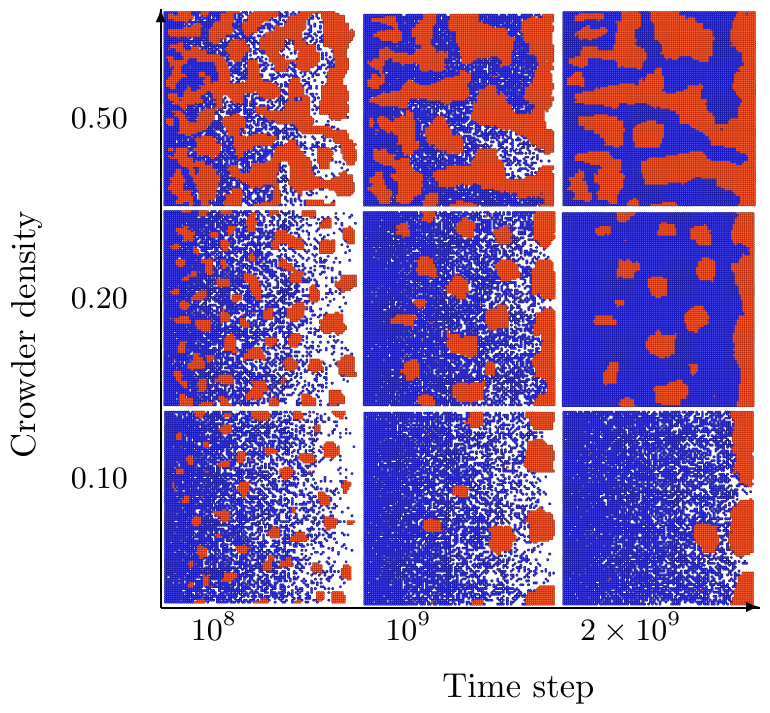}
    \caption{Simulated temporal evolution showing diffusion and pattern formation for 
      \color{black}
      $k_\mathrm{{B}}T_\mathrm{s}/J_\mathrm{int} = 0.25$ 
      \color{black}
      at three distinct crowder densities: 
      top row: $\rho_\mathrm{{C}} = 0.50$;
      middle row: $\rho_\mathrm{{C}} = 0.20$;
      bottom row: $\rho_\mathrm{{C}} = 0.10$, 
      after $10^{8}$ steps (left column), $10^{9}$ steps (middle column), and
      $2 \times 10^{9}$ steps (right column). 
      Crowders are shown in red and tracers in black, with a tracer source on the left side 
      and tracer sink on the right side of the system. The two higher density systems have 
      both arrested with the sink entirely blocked by crowders. 
      The lower density system (bottom row) has not arrested but may do so at longer time scales. Movies showing time evolution of all   
      three systems are included in Supplementary Information~\cite{pre_SI}.}
    \label{fig:temporal evolution}
  \end{centering}
\end{figure}

Temporal evolution of the system is shown in Fig.~\ref{fig:temporal evolution} 
for scaled temperature 
\color{black}
$k_\mathrm{B}{T}_{s}/J_\mathrm{int} = 0.25$
\color{black} 
at crowder density ($\rho_\mathrm{C}$)
values of 0.1, 0.2, and 0.5.
Here, $T_\mathrm{s}$ is the simulation temperature. 
Movies showing the time evolution of all three systems may be viewed in the supplementary information~\cite{pre_SI}.

\color{black} 
Crowders, shown in red, initially aggregate to form clusters which coarsen, drift gradually toward the sink, and aggregate there, 
as shown in Fig.~\ref{fig:temporal evolution}. 
During the initial transient regime, tracer flux changes in response to the evolving microstructure of mobile obstacles. 
Both microstructure and tracer flux eventually reach steady state after $2 \times 10^9$ Monte Carlo steps per lattice site.

\color{black}
Crowder aggregates show a variety of shapes as discussed above;
at higher crowder density they tend to elongate in the direction parallel to the
tracer gradient. Once the crowder clusters reach the sink they
flatten against the surface. At long times, the sink may become fully
covered with crowders. Thermal fluctuations may allow intermittent
tracer diffusion through the surface layer of crowders. However, if the
system becomes filled with tracers entirely, with no remaining vacancies,
tracer diffusion drops to zero and the system remains permanently arrested.

\begin{figure*}
  \begin{centering}
    \includegraphics[width=4.5in]{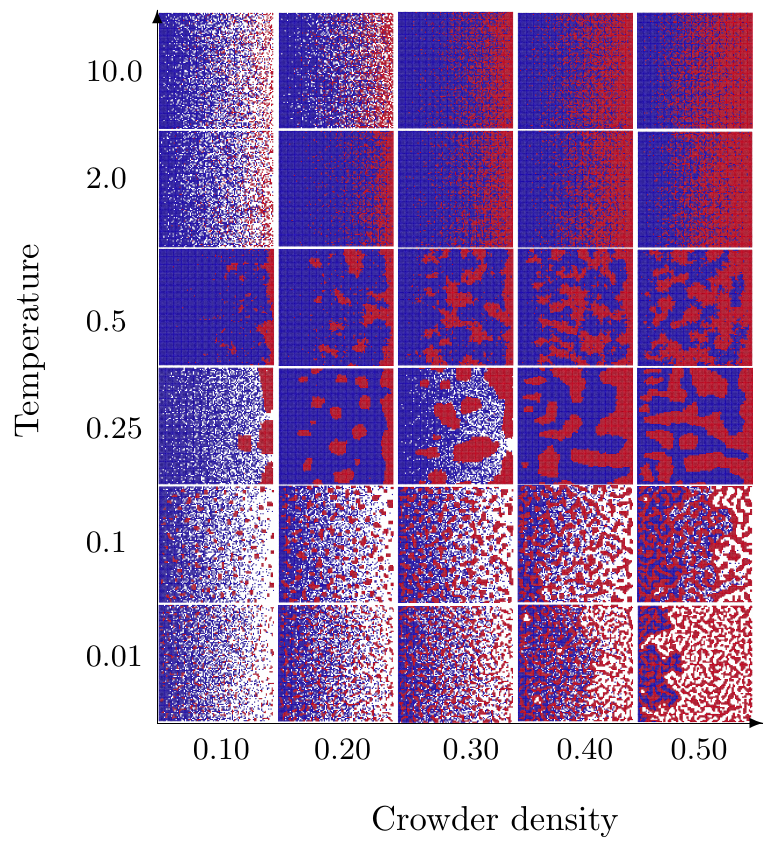}
    \caption{Final configurations
      for varying values of crowder density
      and temperature. The x-axis represents crowder density
      ($\rho_\mathrm{C}$)
      and the y-axis represents
      scaled temperature ($k_\mathrm{B}T_\mathrm{s}/J_\mathrm{int}$).
      Tracers are shown in black and crowders are
      shown in red.}
    \label{fig:equilibrium_phase_space}
  \end{centering}
\end{figure*}

Fig.~\ref{fig:equilibrium_phase_space} shows an overview of the
configurations of the system for a range of crowder densities and
temperatures, after $ 2 \times 10^9$ Monte Carlo steps per lattice
site. Pattern formation depends sensitively on both crowder density
and temperature. We focus on the interesting regimes of the phase
space defined by crowder density, $\rho_\mathrm{C}$, and temperature, $T_\mathrm{s}$, to gain
insight into the resulting microstructure and pattern formation.

At low temperature and low crowder density, crowders form small
aggregates that do not move much or coarsen on the time scale of the
simulation. The resulting microstructure represents a long-lived
metastable state, which is reminiscent of diffusion through fixed
obstables~\cite{nicolau:07}. At low temperature and high crowder
density, crowders quickly aggregate to form a maze-like metastable
microstructure via spinodal decomposition~\cite{cahn:61,ball:90},
through which tracer particles diffuse. At high enough crowder
density, the crowder aggregates percolate and block tracer transport
entirely. Diffusivity near the threshold density for crowder
percolation shows scaling behavior as discussed below.

At low-intermediate temperatures, 
$k_\mathrm{B}T_\mathrm{s}/J_\mathrm{int} = 0.25$,
we observe an intriguing transient behavior. Crowder aggregates form,
coarsen, drift toward the sink, and flatten there. Observed
morphologies also include formation of elongated structures parallel
to the direction of tracer flow.  At high-intermediate temperatures,
$k_\mathrm{B}T_\mathrm{s}/J_\mathrm{int} = 0.5$, 
all crowders drift immediately
to the sink and aggregate there, and tracer diffusion is blocked entirely.

At high temperatures, 
$k_\mathrm{B}T_\mathrm{s}/J_\mathrm{int} \ge 1.0$, 
for low crowder density, crowders aggregate to form a layer covering the sink
but remain sufficiently disordered to allow non-zero tracer
diffusion. At high crowder density, tracer diffusion is fully-blocked.

\subsection*{Dependence of diffusivity on crowder density and temperature}

\begin{figure}
  \begin{centering}
    \includegraphics[]{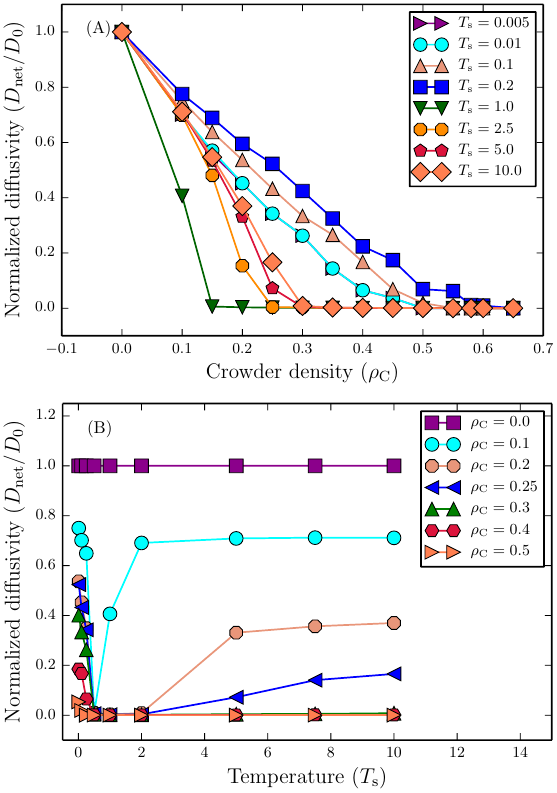}
    \caption{Normalized diffusivity as a function of (A) crowder density ($\rho_\mathrm{C}$)
      at fixed simulation temperatures ($T_\mathrm{s}$), and
      as a function of (B) simulation temperature ($T_\mathrm{s}$)
      at fixed crowder densities ($\rho_\mathrm{C}$).
      Legends show the fixed values of temperatures
      and crowder densities, respectively. 
      \color{black}
      Net tracer flux is computed by counting the rate of tracers annihilating at the sink
      per unit length per unit time.
      We define the net diffusivity of tracers as $D_\mathrm{net}$, and define $D_{0}$ 
      as the net diffusivity of tracers in the absence of crowders. 
      The quantity $D_\mathrm{net}/D_{0}$ is thus calculated as tracer flux (with crowders) 
      normalized by the tracer flux in the absence of crowders.}
    \color{black}
    \label{fig:diffusivity_vs_crowder_density_and_temperature}
  \end{centering}
\end{figure}

Tracer diffusivity as a function of crowder density and temperature is
shown in
Fig.~\ref{fig:diffusivity_vs_crowder_density_and_temperature}.  For
crowder density below 0.3, we observe re-entrant behavior, as tracer
diffusivity first drops to zero and then rises with increasing
temperature. Diffusivity as a function of crowder density also shows
complex, non-monotonic behavior.

To gain insight into these results, we consider various limits where
analytic predictions are possible.

\subsection*{High temperature limit: analytic solution}

In the limit of very high temperature
\color{black}
($k_\mathrm{B}T_\mathrm{s}/J_\mathrm{int} \rightarrow \infty $),
\color{black}
equivalent to $J_\mathrm{int} \rightarrow 0$, crowders have only
excluded volume interactions. In this case, tracers and crowders are
chemically equivalent except for their distinct boundary conditions,
\color{black}
and the system is equivalent to a random walk model with two types of particles, 
where the tracer species has a source and a sink, and the number of crowders is fixed.
\color{black}
In this case, there is no phase separation and we can solve the diffusion equation analytically. The
diffusion equation for this scenario can be written as
\begin{equation}
  \label{eq:diffusion}
  \frac{\partial (T(x) + C(x))}{\partial t} = D \Big( \frac{\partial^2 (T(x) + C(x))}{\partial x^2} \Big)
\end{equation}
where, $T(x)$ and $C(x)$ represent the tracer and crowder density profiles,
position $x$ goes from $0$ (source) to $L$ (sink), and $D$
represents the self-diffusion coefficient of both species.  
\color{black}
To find the steady-state solution of the diffusion equation, as an ansatz we consider a solution of the form


\begin{align}
  & T(x) = a_{0} + a_{1}x + a_{2}x^2 \label{eq:tracer} \\
  & C(x) = b_{0} + b_{1}x + b_{2}x^2 \label{eq:crowder}
\end{align}

and apply the boundary conditions $T(x = 0) = 1$, $T(x = L) = 0$, and $\frac{1}{L}\int_{0}^{L} C(x) dx = \rho_\mathrm{C}$ 
to find the coefficients $\{a_{i}, b_{i}\}$.


The resulting steady-state solution in the high temperature limit is 
\begin{align}
  & T(x) = 1 - (1 - 3 \, \rho_\mathrm{C})\frac{x}{L} - (3 \,
    \rho_\mathrm{C})\frac{x^2}{L^2} \label{eq:tracer1} \\
  & C(x) = (3 \, \rho_\mathrm{C})\frac{x^2}{L^2} \label{eq:crowder1}
\end{align}

\begin{figure*}
  \begin{centering}
    \includegraphics[]{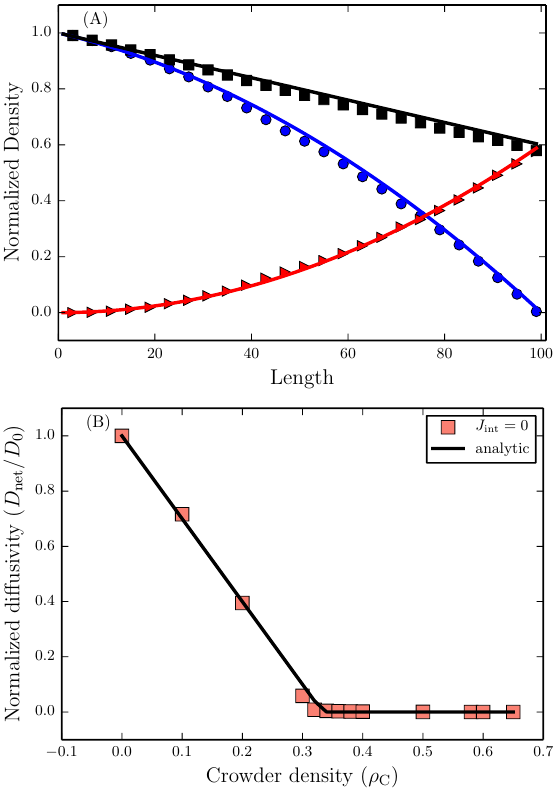}
    \caption{(A) Density profiles of tracers (red triangles), 
      crowders (black circles) and their sum (black squares) 
      for the high temperature limit 
      at a value of crowder density $\rho_\mathrm{C} = 0.2$.
      Solid line represent analytic solutions
      given by Eqs.~\ref{eq:tracer1} -~\ref{eq:crowder1} and their sum.
      (B) Normalized diffusivity as a function of crowder density
      for the scenario where the exchange coefficient of interaction, $J_\mathrm{int} = 0$.
      Squares represent simulated data points and the solid line
      represent analytic solution for diffusivity.}
    \label{fig:density_vs_length}
  \end{centering}
\end{figure*}

\color{black}
Fig.~\ref{fig:density_vs_length} compares this analytical result with Monte Carlo simulation 
data in the case $J_\mathrm{int}=0$, showing time-averaged density profiles for tracer and crowder species; 
analytical and simulation results are in good agreement. Density profiles $T(x)$ and $C(x)$ for tracer and 
crowder species, respectively, were calculated by averaging over the $y$ direction; time averaged over 
the last 10\% of each simulation to capture the steady state behavior; 
and averaged over 10 independent simulations, each lasting at least $2 \times 10^9$ Monte Carlo steps. 
This result shows that even without attractive interactions, crowder density is depleted near 
the tracer source and concentrated near the tracer sink. 

\color{black}

Resulting steady state flux of particles of both types is calculated as $J = -D \frac{d(T(x)+C(x))}{dx}=D (1 - 3 \, \rho_\mathrm{C})$.
Here, $D$ is the self-diffusion coefficient of both tracers and crowders. 
Since crowders cannot exit the simulation box, this flux represents the net flow of tracers. Thus in
the high temperature limit, tracer flux, and thus net diffusivity, is proportional to
$(1 - 3 \, \rho_\mathrm{C})$.  Hence, tracer flux vanishes entirely
for crowder density  $\rho_\mathrm{C} \ge 1/3$. At such high crowder density, after an initial transient, 
the sink is blocked entirely by crowders and tracer flux drops to zero. 
This result is verified via Monte Carlo simulation as shown in Fig. 4(B). 
Here, net diffusivity of tracers is calculated by counting the flux of tracers exiting at the sink, per unit length and time, 
averaged over ten independent simulations, excluding the initial transient before the systems have reached steady state. 
This quantity is normalized by the flux in the absence of crowders.


\color{black}
The good agreement of analytic solution and simulation data in the
limit of $J_\mathrm{int} \rightarrow 0$ case is somewhat intriguing in the light of 
simple exclusion processes in non-equilibrium statistical 
mechanics~\cite{richards:77,privman:05,liggett:13}.
Macroscopic transport principles concerning Eq.~\ref{eq:diffusion}
provides an approximate description of two
species competing for space~\cite{fanelli:10,fanelli:13,galanti:14}.
Nevertheless, it is reasonable to describe the system evolution
in terms of macroscopic transport equation when two species are 
chemically equivalent in the limit of high temperature.

\subsection*{Low temperature limit: scaling behavior near the percolation threshold}

\begin{figure}
  \begin{centering}
    \includegraphics[width=7.5cm]{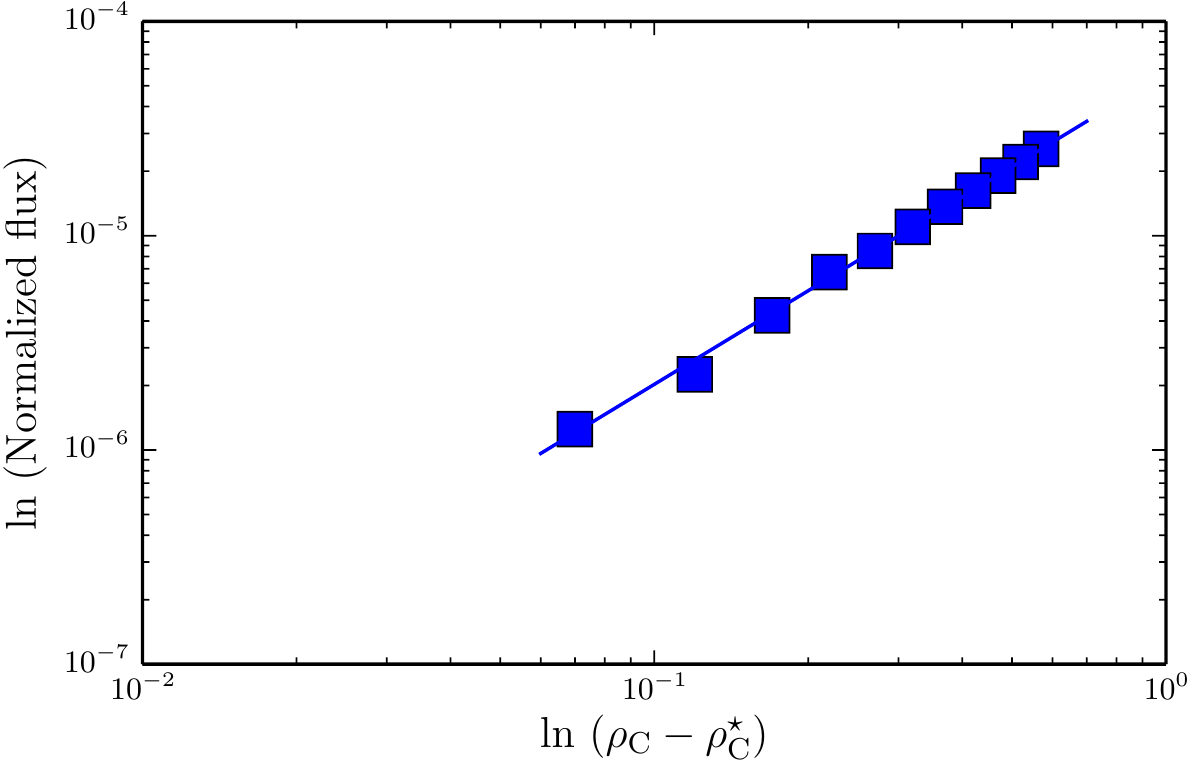}
    \caption{Time-averaged tracer flux as a function of $\left(\rho_\mathrm{C} - \rho_\mathrm{C}^{\star}\right)$
      at a temperature corresponding
      to $k_\mathrm{B}T_{s}/J_\mathrm{int} = 0.1$.    
      Squares represent simulation data and the solid line is
      a linear fit to the simulated data with scaling exponent $\alpha \approx 4/3$.}
    \label{fig:flux_vs_density_fit}
  \end{centering}
\end{figure}

At low temperature, crowders quickly aggregate into clusters that are
essentially immobile and do not drift toward the sink on the time
scale observed in our simulations. At low crowder density, these
clusters act as fixed obstacles and allow continuous tracer diffusion.
At higher density, crowders aggregate into an extended network that
entirely blocks tracer diffusion. Near the threshold density, we
observe scaling behavior. Fig.~\ref{fig:flux_vs_density_fit} shows the dependence
of tracer flux as a function of
$\left(\rho_\mathrm{C} - \rho_\mathrm{C}^{\star}\right)$ on a log-log plot, for
$k_\mathrm{B}T_\mathrm{s}/J_\mathrm{int} = 0.1$. 

Scaling behavior is observed for $\rho_\mathrm{C}^{\star}=0.43$.
The diffusive flux scales as $\left(\rho_\mathrm{C} - \rho_\mathrm{C}^{\star}\right)^\alpha$ with
$\alpha \approx 4/3$. Interestingly, this value of the exponent is in
good agreement with the scaling exponent for conductivity through a random resistor 
network model near its percolation threshold~\cite{fisch:78,derrida:83,pennetta:00}.
Fick's law in steady state for diffusion is equivalent to Kirchoff's law of conductivity through
an electrical network. Hence, steady state diffusion of tracers around randomly distributed obstacles is related to the
electrical conductivity of a random resistor network. This relationship has been previously
noted and explained in the literature~\cite{hofling:13}. 

\color{black}
If the crowders were immobile and randomly distributed with no spatial correlations,
 tracer flux would show scaling behavior near the critical density where open sites form a percolating 
connected pathway between the source and the sink. As the percolation threshold, $p_c$, for the square 
site lattice is $\approx 0.59$, we would expect scaling behavior near $\rho_\mathrm{C}^{\star}= 1 - p_c \approx 0.41$. 
Instead, we found scaling behavior near $\rho_\mathrm{C}^{\star} \approx 0.43$. 
The slight change in the percolation threshold is due to spatial correlations due to crowder interactions~\cite{blumberg:80}.
In the low temperature limit, near the percolation threshold, random walks on a fractal
should produce anomalous diffusion~\cite{havlin:02}. By analogy with the random resistor network, 
we can expect diffusive flux to scale as $(\rho_\mathrm{C} - \rho_\mathrm{C}^{\star})^{\alpha}L^{d-2}$~\cite{redner:12},
where $L$ is the system size.
Thus, in two dimensions, near the percolation threshold, we do not expect the diffusivity to scale in an interesting
way with the system size.
\color{black}

\section*{Discussion}

Although the model described here is remarkably simple, the interaction of phase 
separation with gradient-driven diffusion is nevertheless complex.
The dependence of diffusive tracer flux can be summarized by a single
function of crowder density and temperature only in the high temperature 
limit 
\color{black}
$k_\mathrm{B}T_\mathrm{s}/J_\mathrm{int} \rightarrow \infty$,
\color{black}
where no phase separation occurs. 
The resulting diffusivity in the high temperature limit is
\begin{equation}
  \left.
    \begin{aligned}
      D_\mathrm{net} &= D_{0} (1 - 3\rho_\mathrm{C}) \, \, \,; \, \, \rho_\mathrm{C} < 1/3 \quad\\
      &= 0 \, \, \, ; \, \, \rho_\mathrm{C} \geq 1/3
    \end{aligned}
  \right\},
\end{equation}
where, $D_{0}$ is the diffusivity in the absence of any
crowders in the system.  

In the limit of low temperature near the percolation transition, the system
shows scaling behavior where we find the
following dependence of diffusivity on crowder density:
\begin{equation}
  \left.
    \begin{aligned}
      D_\mathrm{net} &= D_{0} \left(\rho_\mathrm{C} - \rho_\mathrm{C}^{\star}\right)^{4/3} \, \, \,; \, \, \rho_\mathrm{C} < \rho_\mathrm{C}^{\star} \quad\\
      &= 0 \, \, \, ; \, \, \rho_\mathrm{C} \geq \rho_\mathrm{C}^{\star}
    \end{aligned}
  \right\},
\end{equation}
where, $\rho_\mathrm{C}^{\star} = 1 - p_c = 0.43$, is the critical crowder density above which tracer diffusivity drops to zero.

When a multi-component mixture
phase separates in the presence of gradient-driven diffusion, the
resulting tracer flux depends on the evolving
microstructure and in turn influences microstructural evolution. 
Although the multi-species lattice gas model presented in this work is highly idealized, 
it demonstrates this key mechanism. Intracellular transport in complex biological media involves more 
complex interactions than a lattice
gas model can represent. 
In a recent study involving diffusive motion of particles in an environment
of spherical crowders, a non-monotonic dependence of the diffusion rate
on the strength of crowder-diffuser attraction was observed~\cite{putzel:14}.
Relevant experimental studies also include
diffusion in a lipid mixture monolayer at the air-water
interface~\cite{neumann:16}, such as in a Langmuir trough~\cite{gudmand:09}, where
lipid raft formation, may for instance, inhibit gradient-driven diffusion of other molecular species.
\color{black}
In the simplified and highly idealized model presented in this paper, we assume that isolated
``tracers'' and ``crowders'' have the same radius and self-diffusion coefficient. In experimental systems,
two chemical species may of course have completely different properties. Likewise, we have neglected the possibility of 
attractive or repulsive interactions between tracers and crowders, and the potential role of 
hydrodynamic interactions, that is, the scenario in an off-lattice model. 
We plan to address these situations in future work.
\color{black}
 \begin{acknowledgments}
   The authors gratefully acknowledge financial support from
   the National Science Foundation grant DMR-1409658.  
 \end{acknowledgments}

%


\end{document}